\def \yskip{\penalty-50\vskip3pt plus 3pt minus 2pt}
\def \reference{\par \yskip \noindent \hangindent .4in \hangafter 1}
\def \abc#1#2#3#4 {\reference#1, {\sl#2}, {\bf#3}, #4}
\def \blank {\lower 5pt\hbox to 0.75in{\hrulefill}}

\def \s{~\rm{s}}
\def \km{~\rm{km}}

\def \yr{~\rm{yr}}

\documentstyle[11pt,aasms4]{article}

\begin{document}
\small


\title{ 
A MODEL FOR THE OUTER RINGS OF SN 1987A}

\author{
Noam Soker\\
Department of Physics, University of Haifa at Oranim\\
Oranim, Tivon 36006, ISRAEL \\
soker@physics.technion.ac.il }

\begin{center} 
{\bf ABSTRACT}
\end{center} 

I propose a model for the formation of the two outer rings of SN 1987A. 
The main new ingredient is a short-lived, few$\times 10$ years, 
intermediate wind, of velocity $\sim 100 \km \s^{-1}$ and total mass of 
{{{  $\sim 10^{-3} M_\odot -$few$\times 10^{-2} M_\odot$, }}}
which is concentrated near the equatorial plane.
This intermediate wind was formed during the few orbital periods 
when a binary companion entered the envelope of the progenitor of
SN 1987a, when the latter was a red supergiant. 
The intermediate wind formed a hollowed torus with a horseshoe-shaped 
cross section.
The inner regions at the ends of the horseshoe torus are the two outer 
rings of SN 1987A. 
 The other regions of the torus were cleaned by the fast wind blown
after the progenitor turns into a blue giant. 
 The proposed scenario accounts for: ($i$) slow motion of the outer 
rings; ($ii$) the high density contrast of the outer rings to their 
surroundings; ($iii$) the presence of only two outer rings;
($iv$) the displacement of the rings relative to the central star;
($v$) the fact that no planetary nebulae have been 
observed to have a single pair of outer rings. 

{\bf Key words:} supernovae: individual: SN 1987A
$-$planetary nebulae: general
$-$mass loss
$-$star: binaries

\clearpage

\section{INTRODUCTION}

 The nebula around SN 1987A in the Large Magellanic Cloud is dominated
by three rings: an inner small ring and two outer large rings
(Crotts, Kunkel, \& Heathcote 1995; Plait {\it et al.} 1995;
Panagia {\it et al.} 1996).
 The inner bright ring has a diameter of $0^{\prime \prime}.81$ 
and a semiminor axis of $0^{\prime \prime}.57$ (Burrows {\it et al.} 1995).  
 The two outer rings are not identical, with the semimajor and semiminor
axes of the northern (southern) ring being $1^{\prime \prime}.77$ 
($1^{\prime \prime}.84$) and $1^{\prime \prime}.30$ ($1^{\prime \prime}.58$), 
respectively (Burrows {\it et al.} 1995).  
  There are several interesting properties of the rings. 
  (a) The center of the line joining the centers of the two outer rings 
      is displaced by $0^{\prime \prime}.4$ from the central star, 
      in the north-west direction (Burrows {\it et al.} 1995).  
      This is a displacement of $\sim 10 \%$ of the size of the outer rings.
  (b) A very careful study of the inner ring suggests that its center
      is displaced in the same direction, but only by 
      $\lesssim 0^{\prime \prime}.03$, i.e., $\lesssim 2 \%$ displacment. 
  (c) There is a huge density contrast (a factor of $\sim 100$) between 
      the outer rings and their surroundings (Burrows {\it et al.} 1995;
      Panagia {\it et al.} 1996).
  (d) The two outer rings are complete rings, and there are no other pairs
      of rings observed in the system. 
  (e) The expansion velocity of the outer ring is 
      small (Burrows {\it et al.} 1995; Panagia {\it et al.} 1986).  
  
As noted by Burrows {\it et al.} (1995), these properties of the 
rings impose severe difficulties on models for the formation of the rings.  
 Burrows {\it et al.} (1995) find problems in the interacting winds 
model (e.g., Blondin \& Lundqvist 1993; 
{{{   Lloyd, O'Brien \& Kahn 1995; Martin \& Arnett 1995}}}), 
models based on jets,
the interacting wind from a companion model proposed by  
Podsiadlowski, Fabian, \& Stevens (1991), and a few other scenarios. 
 {{{  In the standard interacting winds model, the two outer rings
are considered to be the limb-brightened surface of two lobes,
one at each side of the equatorial plane. 
 Such a structure was also suggested by Crotts {\it et al.} (1995),
based on their analysis of light echoes. 
 However, the huge density contrast between the rings and their 
 surroundings is problematic in standard interacting winds models
(e.g., Lloyd {\it et al.} 1995; Martin \& Arnett 1995). 
 The simulated emission-measure image obtained by Martin \& Arnett (1995;
their fig. 5) does not produce two bright rings. 
Instead, only two arcs are seen in high contrast to
the surroundings. }}}

 Meyer (1997) further criticizes models based on the interaction of the 
fast blue supergiant (BSG) wind with the dense and slow red supergiant
(RSG) wind. 
 Instead, Meyer (1997) proposes that the rings are formed from 
ionization that induces hydrodynamic motions. 
 I find the density contrast between the rings and their surroundings 
to be too low in Meyer's model.  
 In addition, the inner ring would prevent the ionization of the equatorial 
region between the two outer rings, further reducing the efficiency of 
the model and the density contrast. 
 
 To the properties of the outer rings listed above, 
I would add that a nebula composed 
of a single pair of outer rings is not observed in planetary nebulae (PNs).
{{{  The closest structure is that of the Hourglass nebula
(also named MyCn 18 and PN G 307.5-04.9),
which is a bipolar PN (Sahai \& Trauger 1996).
 I find three properties that significantly distinguish between 
the structure of the rings of the Hourglass nebula and the two outer rings 
of SN 1987A: ($i$) There are several rings in each of the two lobes of the
Hourglass nebula, as opposed to one ring on each side of the equatorial plane
of SN 1987A; ($ii$) There is a rich structure between and around the
rings of the Hourglass nebula, as opposed to the large density contrast
between the outer rings of SN 1987A and their surroundings;
and ($iii$) There are large variations in the brightness along each of 
the many rings of the Hourglass nebula, much larger than the
variation along the two outer rings of SN 1987A (not considering 
short scale fluctuations). 
 The structure of the Hourglass nebula clearly show two lobes, as predicted
by the interacting winds model.
In the present paper I propose a different model, in which the two 
outer rings of SN 1987A are {\it not} part of two lobes. }}}
 Instead, I propose a different model ($\S 2$), which incorporates a 
short-lived intermediate wind caused by the onset of a common envelope 
($\S 3$).  
 The discussion and summary are in $\S 4$.

\section{THE PROPOSED MODEL}

 I start by describing a 2D numerical simulation performed as a 
speculative effect for shaping proto-PNs (Soker 1989).
 In that paper I assumed that in the transition from the asymptotic giant
branch (AGB) to the PN, the star has a short mass loss episode, a ``pulse'', 
due to an interaction with a binary companion.  
 In this pulse, mass loss occurs close to the equatorial plane, and at a 
velocity faster than that of the slow wind, but slower than the later fast 
wind. 
 In that specific simulation I assumed that the pulse occurs 600 
years after the end of the slow wind, it lasts for 50 years, and it is 
concentrated within an angle of $\sim 10 ^\circ$ from the 
equatorial plane.
 The velocity of the material in the pulse is $200 \km \s^{-1}$,
and the total mass $3 \times 10^{-4} M_\odot$.
 This pulse runs into the slow wind, which has a density contrast of 
6 between the equator (high density) and polar directions, 
a total mass loss rate of $10^{-5}  M_\odot \yr^{-1}$, and
a velocity of $10 \km \s^{-1}$.
 As the pulse hits the slow wind, a high pressure region is formed in 
a small region in the equatorial plane (in 3D it has the shape of a ring).
The fast release of energy resembles an explosion, and it creates a shell
expanding from the high pressure region.  
 Since there is no slow wind material inward, the shell has the shape
of a horseshoe as observed in the symmetry plane 
(the plane perpendicular to the equatorial plane). 
 A schematic drawing of a horseshoe shape is seen in phases 3 and 4
of Figure 1.  
   For SN 1987A different parameters should be used, so the results of
Soker (1989) should be applied qualitatively rather than quantitatively. 
 It should be emphasized that the qualitative result, of forming a 
horseshoe-shaped torus, does not depend on a density contrast between the 
equator and poles. 
 The inner regions of the horseshoe (i.e, the regions at the ends of
the horseshoe above and below the equatorial plain)
are
(a) denser than most of the other regions of the torus, 
(b) their velocity is the lowest, and
(c) they are extended in a radial direction more or less. 

 Assume now that the central star starts to blow a fast wind, which
hits the horseshoe torus and accelerates it. 
 The inner regions of the horseshoe torus will have the lowest acceleration
since they are denser, and moreover, they are elongated in a more or less
radial direction (phase 4 of fig. 1).
  Therefore, there will be a time in the evolution of this flow when
most of the original torus has already been expelled to large distances 
from the star by the fast wind, and hence it has low density. 
 The original inner regions of the torus, by contrast, will expand slower, 
be relatively close to the star, and be dense, much denser than the 
fast wind material around it.
{{{  The density contrast between the inner regions of the horseshoe 
torus and their surrounding, will be much larger than that in the numerical
simulation of Soker (1989), since the fast wind blown by the progenitor,
will compress the rings, and in addition will clean their surroundings. }}}
 The dense regions now form two rings, one at each side of the 
equatorial plane. 
 
 Let me try to sketch a scenario for the formation of the three
rings around SN 1987A based on the discussion above.
 One assumption that goes here and in the explanation for the displacement 
of the rings from the exploding star is that of a binary companion. 
 There are other reasons to support a binary companion. 
 First, Chevalier \& Soker (1989) show that a deformed envelope due to fast 
rotation can explain the asymmetrical explosion of SN 1987A, which is inferred 
from polarization data. 
 The direction of polarization is in the direction of the symmetry axis 
of the inner ring (or perpendicular to it, depending on the time and 
emission lines). 
 The angular velocity required can be gained from a companion of mass
$\gtrsim 0.5 M_\odot$. 
 Second, the merger of a $\gtrsim 3 M_\odot$ secondary with the progenitor's 
envelope makes the envelope shrink, hence the transition of
the progenitor to a blue star (Podsiadlowski, Joss, \& Rappaport 1990). 
 {{{   Third, Collins {\it et al.} (1998) argue that a merger with a
companion of mass $M_s \gtrsim 0.5 M_\odot$ will lead to the formation
of the inner ring of SN1987A. }}}

 In a preliminary study (Soker 1998) I considered two possible tracks to 
the proposed evolutionary sequence. 
 In only one of these tracks, the one described here, I find 
a mechanism to form the short-lived intermediate wind ($\S 3$). 
 The different stages of the proposed scenario are depicted in Figure 1. 
\newline  
  {\bf Phase 1:} Slow wind from the RSG progenitor.
 It may have a higher mass loss rate in the equatorial plane due to
slow rotation or a tidal interaction with the companion.
If the companion is in an eccentric orbit,
{{{  and its tidal effect on the progenitor influence the mass loss
rate, then }}}
 this wind may be displaced 
relative to the central star and perpendicular to the symmetry axis 
(Soker, Rappaport, \& Harpaz 1998). 
 This may explain the large displacement of the outer rings of 
SN 1987A from the central star. 
{{{  This process requires high eccentricity, and that the 
companion be close enough to significantly alter the mass loss rate from 
the progenitor as it moves along its orbit
 (in any case the companion must be close enough before entering the 
common envelope). }}}
\newline
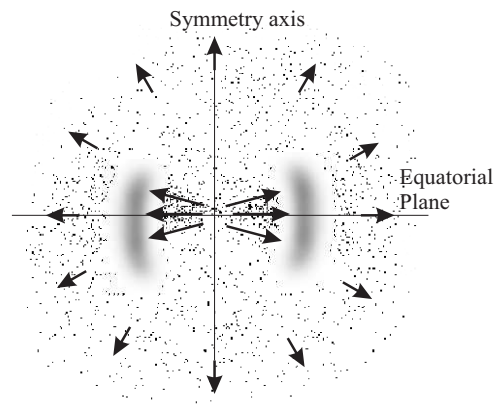
{\bf Phase 2:} The binary system blows a faster wind concentrated in the 
equatorial plane. I term this wind ``intermidiate wind''.  
A horseshoe-shaped torus is formed. 
 In the next section I suggest that such a wind occurs as the companion 
enters the envelope of the primary (the progenitor of 1987A).
\newline
{\bf Phase 3:} A slow dense wind is blown mainly in the equatorial plane. 
 The inner edge of this wind forms the inner ring of SN 1987A.
 The slow wind concentration in the equatorial plane is due to the
faster rotation of the envelope {{{   (Collins {\it et al.} 1998), }}} 
and possibly by excitation of nonradial pulsation by the secondary 
orbital motion inside the primary's envelope (Soker 1993). 
 This lasts several thousand years (Podsiadlowski {\it et al.} 1990). 
 A dense ring in the equatorial plane is seen in several PNs 
(e.g., NGC 2346; Sp 1) known to have a close binary companion at their 
centers (Bond \& Livio 1990), 
{{{  and in VY CMa, an M supergiant which
will end as a supernova (Kastner \& Weintraub 1998). }}}
\newline
{\bf Phase 4:} Several thousand years after the secondary enters the 
envelope the primary shrinks to a BSG (Podsiadlowski {\it et al.} 1990), 
and blows a fast wind.  
 This wind pushes most of the previous dense wind to large distances, beside
the very dense regions which are also elongated in radial directions: 
the inner regions of the horseshoe torus and the dense disk in the 
equatorial plane. 
 The contraction of the rotating progenitor to a BSG spins-up the envelope.
 This can lead to the concentration of the BSG fast wind toward the
equatorial plane. 
\newline 
{\bf Phase 5:} Just before explosion the system contains three rings:
two outer rings, one at each side of the equatorial plane, which 
are the remnant of the inner regions of the horseshoe torus,
and an inner dense ring in the equatorial plane. 
 
\section{THE INTERMEDIATE WIND}

 The three-dimensional numerical simulations performed by Livio \& Soker 
(1988; see also Terman, Taam \& Hernquist 1994, and references therein) 
show that the onset of a common envelope phase, i.e., 
the entrance of the secondary star to the primary's extended envelope, 
results in a high velocity flow in the equatorial plane.
 This is the mechanism for the intermediate wind (phase 2 in fig. 1). 
 The primary mass in the simulations of Livio \& Soker (1988) was
$5 M_\odot$ and the secondary mass was $M_s=0.3 M_\odot$ or
$M_s = 1.4 M_\odot$.
The calculations started with nonrotating envelope and
with the secondary outside the envelope.  
 Most of the envelope matter in this equatorial high velocity flow
has a velocity of $v_f < 2 v_{\rm Kep}$, where $v_{\rm Kep}$ is the Keplerian
velocity on the surface.
However, small fraction of this flow has a velocity of $\sim 3 v_{\rm Kep}$,
especially in the case with the high secondary mass. 
 For the progenitor of SN 1987A at the onset of the common envelope 
(Podsiadlowski {\it et al.} 1990) the surface Keplerian velocity 
is $v_{\rm Kep} \simeq 60 \km \s^{-1}$, and a small fraction
of the flow will reach a velocity of $\gtrsim 100 \km \s^{-1}$. 
 Two processes accelerate matter to high speed, gravitational scattering 
and the high pressure formed near the secondary. 
 Numerical simulations are needed to determine the amount of mass 
in this flow which has high velocity. 

 The numerical results were further discussed by Soker (1995). 
 Let the envelope surface near the equatorial plane rotate at velocity 
$v_{\rm rot} = \delta v_{\rm Kep}$.  
 An envelope particle scattered through an angle $\theta$ by the secondary 
gravitational field will have a velocity $v_f$ relative to the primary's
core given by (Soker 1995) 
\begin{eqnarray}
v_f = v_{\rm Kep} [\delta^2+2(1-\delta)^2(1-\cos \theta)-2^{3/2}
\delta (1-\delta)(1-\cos \theta)^{1/2} \cos \theta ]^{1/2}.
\end{eqnarray}
 Maximum velocity, $v_{\rm max} =v_{\rm Kep} (2-\delta)$, 
is obtained by particles scattered in the direction of the
envelope rotation, $\theta = 180^{\circ}$.
 The high pressure around the secondary can further increase the final 
velocity, as was obtain in the numerical simulations. 

 The assumption of free particle scattering is reasonable as long as 
most of the envelope material influenced by the secondary is on one 
side and near the equatorial plane. 
 This will be the case when the secondary is not deep in the envelope.
 If not, then matter from the two sides will collide on the accretion line
as in the classical Bondi-Hoyle accretion flow. 
 The velocity along the accretion line is less than the escape velocity
from the primary surface. 
 The one-side condition also requires that the accretion radius of the 
secondary $R_A$ is not too small relative to the density scale height in 
the envelope $H$, $R_A \gtrsim H$.
 The accretion radius is given by 
$R_A \simeq 2 G M_s / v^2_{\rm rel}$, where 
$v_{\rm rel} = v_{\rm Kep} (1-\delta)$ is the relative velocity of the
secondary and envelope, assuming $\delta \lesssim 0.8$ (otherwise the 
sound speed in the envelope should be included). 
 The density scale height in the outer envelopes of red giants
is $H \sim 0.3R_\ast-0.5 R_\ast$ where $R_\ast$ is the stellar radius. 
For $M_s \ll M_\ast$, where $M_\ast$ is the primary mass,
the Keplerian velocity on the stellar surface 
is $v_{\rm Kep} = ( G M_\ast / R_\ast )^{1/2}$.
 The condition $R_A \gtrsim H$ becomes 
\begin{eqnarray}
M_s \gtrsim 0.2 (1-\delta)^2 M_\ast.
\end{eqnarray}
 
 The second condition for having a fast flow is that the 
envelope rotate much below the Keplerian velocity when the 
secondary enters the envelope, $\delta \lesssim0.5$.
 As the giant primary expands it loses mass, a process which
tends to increase the orbital separation.
 On the other hand, tidal forces decrease the orbital separation. 
 The slow envelope rotation condition means that there is no 
synchronization between the orbital rotation and envelope rotation
when the secondary enters the primary's envelope.
 This occurs if the secondary cannot bring the primary to corotate,
and ``tidal catastrophe'' causes the secondary to spiral inward fast. 
 The condition for tidal catastrophe to develop from synchronized 
orbital motion is (Darwin 1879)
$I_{\rm env} > I_s/3$, where $I_{\rm env} = \eta M_{\rm env} R^2_\ast$
is the envelope's moment of inertia with $\eta \sim 0.2$ for giants,
$I_s = M_s a^2$ is the secondary moment of inertia,
$M_{\rm env}$ is the envelope mass, and $a$ is the orbital separation.
For synchronized orbital motion $\delta = (R_\ast/a)^{3/2}$. 
 As the secondary spirals in during the tidal catastrophe it further
 spins-up the envelope. 
 Taking this into account, the condition $\delta \lesssim 0.5$ 
requires the tidal catastrophe to occur when $a \gtrsim 2.5 R_\ast$. 
 Substituting $I_{\rm env} = I_s/3$ at $a=2.5 R_\ast$, and using the
expressions for $I_{\rm env}$ and $I_s/3$, 
the condition for the envelope not to rotate too fast becomes 
\begin{eqnarray}
M_s \lesssim 0.5 \eta M_{\rm env} \sim 0.1 M_{\rm env}.
\end{eqnarray}
 From the two conditions given by equations (2) and (3) we find,
after substituting $\delta=0.5$ in equation (2),
that the condition on the primary for having a fast equatorial flow at 
the onset of a common envelope is 
\begin{eqnarray}
M_{\rm env} \gtrsim 0.5 M_\ast.
\end{eqnarray}
 The condition on the core mass is 
$M_{\rm core} = M_\ast-M_{\rm env} \lesssim M_{\rm env}$.
Of course, the high flow requires the secondary to have the mass in
the right range.
  For SN 1987A progenitor this range according to the estimate above is
$0.7 M_\odot \lesssim M_s \lesssim 1.5 M_\odot$. 
 However, considering the crude discussion above, a factor of
2 to this range can reasonably be assumed here 
$0.5 M_\odot \lesssim M_s \lesssim 3 M_\odot$. 
 This is more true for the upper limit, since the pressure around massive
secondaries can further accelerate material in the flow, hence allowing 
larger value of $\delta$, which means the tidal catastrophe can occur
when the secondary is closer than $2.5 R_\ast$ to the primary. 
{{{   As mentioned in $\S 2$, merger with a companion of mass 
$M_s \gtrsim 0.5 M_\odot$ is required also for the asymmetrical explosion
of SN1987A in the model of Chevalier \& Soker (1989), and for the
formation of the inner ring in the model of Collins {\it et al.} (1998). }}}

 Condition (4) is clearly met for the progenitor of SN 1987A, which had
a core of $\lesssim 6 M_\odot$ and an envelope of 
$M_{\rm env} \gtrsim 10 M_\odot$ at the onset of the common envelope.
 After the onset of the common envelope, and until the explosion of SN 1987A,
relatively little mass was lost by the progenitor.
 Therefore, the denser inner region of the horseshoe torus that was formed
by this fast equatorial wind (the intermediate wind) survived until
the explosion. 
  As I now show, this does not happen in PNs. 
 The masses of central stars of PNs are $\gtrsim 0.6 M_\odot$. 
For condition (4) to be fulfilled, the envelope mass at the onset
of a common envelope in a progenitor of a PN must be
$M_{\rm env} \gtrsim 0.6 M_\odot$. 
 The onset of the common envelope releases a little mass relative to the
envelope mass. 
 Unlike the progenitor of SN 1978A which retained most of the envelope by 
the time of the explosion, a progenitor of a PN loses {\it almost all} 
of its envelope before turning to a PN.  
 This massive outflow will erase the horseshoe-torus by the time the
star turns to a PN.  
 Therefore, even if a horseshoe torus is formed during the AGB phase of the
progenitor of a PN, I would not expect to observe a prominent signature
of it by the PN phase.  
  Some imprints of it may however be present on the outskirts of the main
nebular shell, i.e., some complicated structures on the outer shell of
PNs whose progenitors have gone through a common envelope
phase may have resulted from the onset of the common envelope. 
  The shell inner to these imprints should have a mass of 
$\gtrsim 0.6 M_\odot$. 
 It is hard to tell what these imprints look like without numerical
simulations, but complicated structures are seen on the outer shells of 
several PNs with concentration of mass in the equatorial plane, 
e.g., A14, JnEr 1, and NGC 6853 
(see images in, e.g.,  Manchado {\it et al.} 1996).
  
The intermediate wind will lead to a deviation from axisymmetry. 
This is because the phase of the onset of the common envelope lasts
for only a few orbital periods. 
 Therefore, the mass loss rate and velocity of the intermediate wind
will change significantly with the secondary location during each 
orbital period (Terman {\it et al.} 1994).
 This may explain the displacement of the outer rings of SN 1987A relative
to the central star. 
{{{  It is hard to predict the displacement without performing
numerical simulations, since the effect of the intermediate wind in
forming the horseshoe torus is due mainly to the kinetic energy
of the intermediate wind, rather than its momentum. }}}
 Other possible mechanisms for displacement are discussed in the next section.

\section{DISCUSSION AND SUMMARY}

 The main new idea incorporated into the proposed model for the formation of
the three rings of SN 197A is a short-lived, few$\times 10$ years, 
intermediate wind, of velocity $\sim 100 \km \s^{-1}$ and a total mass of 
{{{  $\sim 10^{-3} M_\odot -$few$\times 10^{-2} M_\odot$, }}}
which is concentrated near the equatorial plane (phase 2 in fig. 1). 
 The proposed scenario accounts for: ($i$) slow motion of the outer 
rings; ($ii$) the high density contrast of the outer rings to their 
surroundings; ($iii$) the presence of only two outer rings;
and it may account for ($iv$) the displacement of the three rings 
relative to the central star, via eccentric orbit or a short duration
for the onset of the common envelope; 
($v$) the fact that no PNs have been observed to have a single pair
of outer rings ($\S 1$; $\S 3$)

 The displacement deserves some discussion. 
  All three rings are displaced in the same direction relative to the 
exploding star, though the displacement of the inner ring is very small. 
 The general direction of the displacement is north-west. 
 Several processes can cause displacement of circumstellar 
nebulae: {\bf (a)} Interaction with the ISM: 
 Interaction of SN 1987A progenitor's wind with the ISM has been proposed 
(Wang \& Wampler 1992; Wang, Dyson \& Kahn 1993) to explain the 
``Napoleon's Hat'' (e.g., Wampler {\it et al.} 1990), the diffuse
nebulosity on the northern side, extending to a distance of 
$\sim 5 ^{\prime \prime}$ from the exploding star. 
 Similar structures, of a diffuse nebulosity extended from the central star
to a one-sided bow filament,  are seen in interacting PNs, e.g.,
IC 4593 (Corradi {\it et al.} 1997; Zucker \& Soker 1993).
 This interaction, though, cannot explain the displacement of the rings
since the displacement is toward the north, while the ISM is supposed to
flow from north to south relative to SN 1987A 
{{{  (see also Martin \& Arnett 1995). }}}
 In addition, such an interaction would have destroyed the complete 
elliptical shape of the rings. 
 The ISM, therefore, is not the cause of the rings' displacement relative
to the exploding star.
{\bf (b)} Interaction with a wide companion, having an orbital 
period of several$\times 10^4 \yr$ (Soker 1994). 
Such a companion has not been found around SN 1987A. 
{\bf (c)} An eccentric close companion (Soker {\it et al.} 1998). 
As discussed in previous sections, this may have cause the mass loss
from the primary progenitor prior to the common envelope to have
a non-axisymmetrical wind if the orbit was eccentric.   
{\bf (d)} In $\S 3$ I propose another mechanism, 
the onset of the common envelope itself: 
if the intermediate wind is blown in only a few orbital periods.

 According to the proposed model, the matter in the outer regions 
of the horseshoe torus was expelled outward to the two outer rings 
in the radial direction, and to large distances between the rings 
(phase 5 in fig. 1).
 The location and density of this matter at present depends on its 
interaction with the ISM.  
  It is quite possible, according to the proposed model, that the 
``Napoleon's Hat'' is {{{  related to }}} 
the extended matter between the rings (outer regions of the horseshoe 
torus), which was compressed by the ISM in the north. 
 The outer regions of the horseshoe torus in the other directions may have
been dragged to very large distances by the flowing ISM. 
{{{   If instabilities form dense blobs, then these blobs will
be expelled less efficiently, and maybe located closer to the two outer 
rings (but still further away from the exploding star). }}}
 {{{  As with other models, based on the standard interacting winds
model, the interaction between the fast wind and the inner
equatorial ring will result in material flowing between the two rings.
 This material will have densities much below the density of the outer rings,
though. 
 Therefore, the presence of low density gas between the two outer rings does
not contradicts the newly proposed model. }}}

{\bf ACKNOWLEDGMENTS:} 
{{{  I thank the referee, Philipp Podsiadlowski, for detailed and useful
comments. }}} 
I thank Saul Rappaport for several helpful comments.
 This research was in part supported by a grant from the 
Israel Science Foundation. 


$$
$$

$$
$$

$$
$$

\noindent {\bf Figure 1:} Schematic illustration of the proposed scenario
for the formation of the three rings of SN 1987A.
 The horseshoe shape is the projection of a 3D hollowed torus 
on the symmetry plane. 

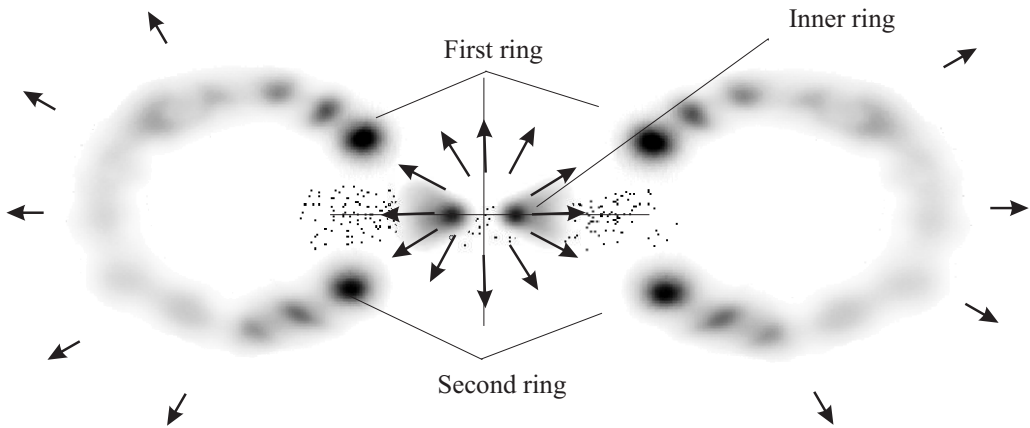
\end{document}